\documentclass{article}

\usepackage[final]{neurips_2019_ml4ad}

\usepackage[utf8]{inputenc} 
\usepackage[T1]{fontenc}    
\usepackage{hyperref}       
\usepackage{url}            
\usepackage{booktabs}       
\usepackage{amsfonts}       
\usepackage{nicefrac}       
\usepackage{microtype}      
\usepackage{graphicx}
\DeclareGraphicsExtensions{.pdf,.png,.jpg}

\title{Self-Driving like a Human driver instead of a Robocar: Personalized comfortable driving experience for autonomous vehicles}

\author{I. Bae, J. Moon, J. Jhung, H. Suk, T. Kim, H. Park, J. Cha, J. Kim, D. Kim and Shiho Kim\\
Seamless Transportation Lab (STL)\\
Yonsei University, Incheon, 21983, Korea\\
\texttt{shiho@yonsei.ac.kr}
}

\begin{document}

\maketitle

\begin{abstract}
This paper issues an integrated control system of self-driving autonomous vehicles based on the personal driving preference to provide personalized comfortable driving experience to autonomous vehicle users. We propose an Occupant’s Preference Metric (OPM) which is defining a preferred lateral and longitudinal acceleration region with maximum allowable jerk for users. Moreover, we propose a vehicle controller based on control parameters enabling integrated lateral and longitudinal control via preference-aware maneuvering of autonomous vehicles. The proposed system not only provides the criteria for the occupant’s driving preference, but also provides a personalized autonomous self-driving style like a human driver instead of a Robocar. The simulation and experimental results demonstrated that the proposed system can maneuver the self-driving vehicle like a human driver by tracking the specified criterion of admissible acceleration and jerk.
\end{abstract}

\section{Introduction}

For fully automated self-driving vehicles, the driver is no longer required to engage in the driving task, hence, the driver becomes a passive passenger or occupant, however, it may cause undesirable motion sickness. Since the motion sickness is a condition of the conflict between visually perceived movement and the vestibular system's sense of movement [1-2]. The human organs of balance are in essence biological accelerometers and this means that they are sensitive to accelerations. The vehicle’s acceleration and jerk, which is the time derivatives of the acceleration, can significantly impact on occupant’s comfort. Careful and appropriate control of the acceleration, braking, and steering action of the vehicle can improve the comfort of the occupants. We define an Occupant Preference Metric (OPM) based on the significant parameters of a vehicle motion, which defining a preferred lateral and longitudinal accretion region with maximum allowable jerk for autonomous vehicles, from smooth driving to aggressive driving. Furthermore, we proposed an integrated control system for enabling human driver-like maneuvering of autonomous vehicles to provide a personalized comfortable driving experience to users. The proposed overall control system and whole planning strategies are implemented into a real autonomous vehicle to validate the proposed OPM-aware control system via experiments.

\section{Modeling Occupant’s Preference Metric}

Through the review of the previous studies which is related with the comfort criterion of passengers and drivers, we designed the Occupant’s Preference Metric (OPM) defining a preferred lateral and longitudinal acceleration region with maximum allowable jerk. It is employed the five significant parameters which are directly related with a vehicle motions as follow,

\begin{equation}
OPM=\left\{{a_{(+)x}}_{opm},{a_{(-)x}}_{opm},\left|{a_y}_{opm}\right|,\left|{z_x}_{opm}\right|,\left|{z_y}_{opm}\right|\right\}
\end{equation}

where the ${a_{(+)x}}_{opm}$ is the occupant’s discomfort threshold of the longitudinal acceleration, ${a_{(-)x}}_{opm}$ is the longitudinal deceleration threshold, is the lateral acceleration threshold, $|{z_x}_{opm}|and |{z_y}_{opm}|$ is the maximum allowable longitudinal and lateral jerks.

A large acceleration or jerk can make the passenger discomfort even the short periods of time. For example, in the comfort criterion of the public transportation, which is presented with blue region in Figure 1, the average freestanding of passenger in the subway measured within a constant acceleration of $0.93 m/s^2$, and largest jerk average presented about $0.6 m/s^3$ [3]. When those values get too high the passenger in a moment, it is difficult to maintain their postures.

\begin{figure}
  \centering
  \includegraphics[width=1.0\textwidth]{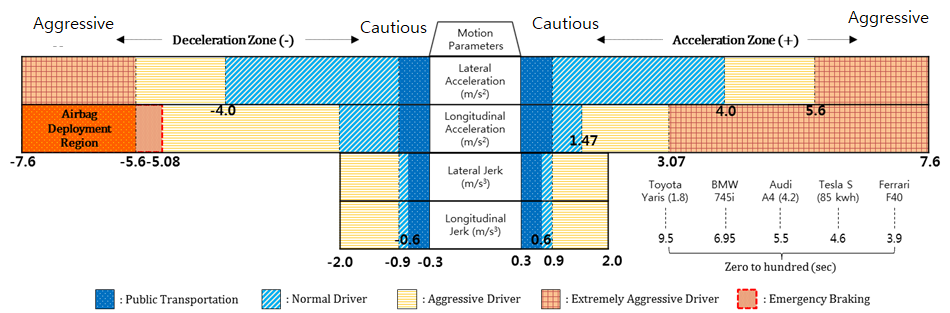}
  \caption{Proposed occupant’s preference metric (OPM).}
\end{figure}

\begin{figure}
  \centering
  \includegraphics[scale=0.35]{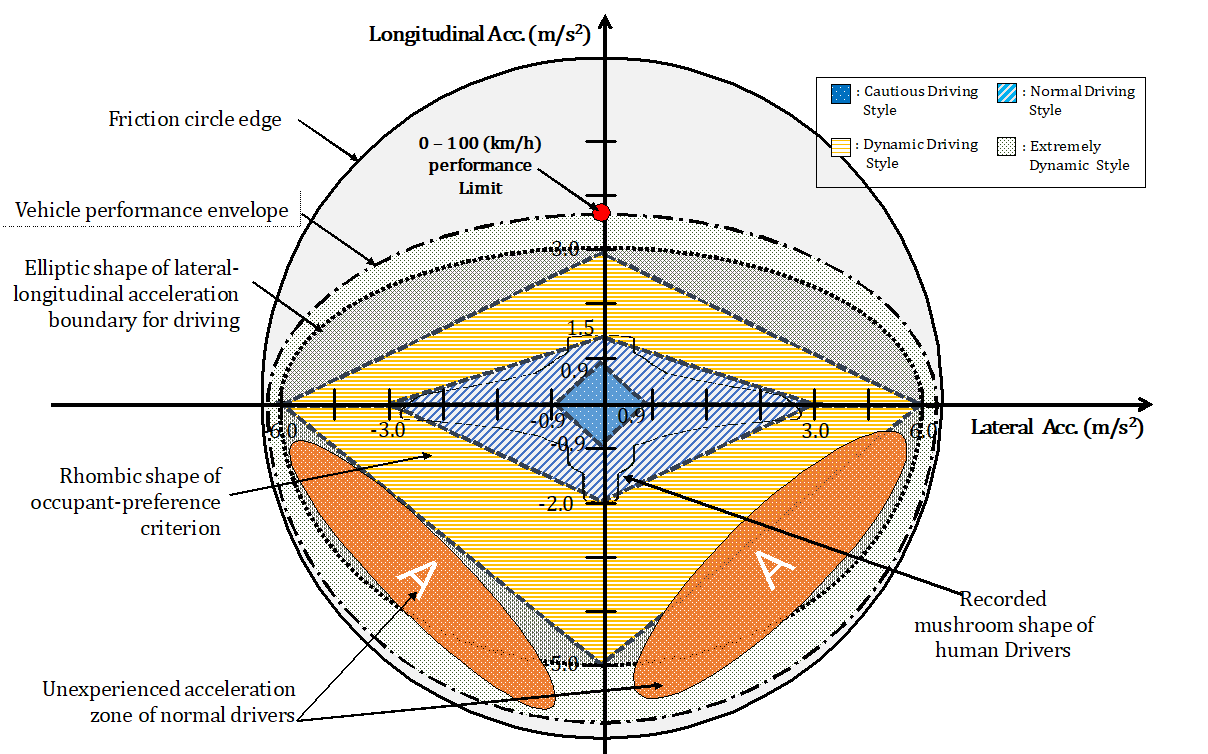}
  \caption{Occupant’s preference metric (OPM) displayed on the G–G diagram along with friction circle and performance envelop of the vehicle. Note that the units of the x- and y-axes are in m/s2, instead of acceleration of gravity units.}
\end{figure}

For the longitudinal acceleration, the discomfort threshold presented around $0.93m/s^2$, ranging up to $1.47m/s^2$. The jerk threshold for discomfort presented about $0.3m/s^3$, ranging up to $0.9m/s^3$. An autonomous car only carries a seated occupant, normally, accelerations might be reasonable to set the limit value around $2m/s^2$ and jerk to $0.9m/s^3$. Particularly, in vehicle deceleration, human driving tests revealed that $99\%$ of braking maneuvers are approximately ranging from $-0.5m/s^2$ to $-2m/s^2$. Moon et. al categorized the braking comfort region within $-2m/s^2$ based on the human manual driving test for design of cruise control deceleration algorithm [4]. In addition, Moon’s human driving data shows that the maximum deceleration and the maximum acceleration values were measured in the test are $-5.08m/s^2$ and $3.07m/s^2$, respectively. These thresholds of the longitudinal acceleration and deceleration values could be overlapped with the boundaries of G-G diagram of Bosetti’s [5]. Moreover, they reported that the jerk is also limited, typically, it does not exceed $2m/s^3$, except emergency braking.

According to [6], the longitudinal accelerations are combined with much higher lateral accelerations, up to about $4m/s^2$. And, it was also recommended the lateral and the longitudinal acceleration thresholds related to nominal driver preference are $0.4g$ and $0.2g$ in case of both passenger car and heavy truck. Maximum acceleration is estimated around $-5.1m/s^2$ to prevent an emergency situation [1] in emergency braking region of Fig. 1. As above mentioned the previous human driving studies, we can classify the four-representative driving type such as the public transportation, normal, aggressive, and extremely aggressive style as shown in Fig. 1. 

The basic approach of the proposed control system is that the driving motion of the autonomous vehicle must below the OPM threshold values as a preference setting of occupants which can be acceptable of comfort criteria for the personalized driving experience. The lateral and longitudinal acceleration profiles displayed on G-G diagram represents the combined results of driver’s preferences and perceived risk level corresponding to the dynamic motion at a given environment. Racing drivers manipulate the vehicular dynamics almost up to the end-range of the performance envelop, whereas normal drivers cannot utilize the full available capability of the vehicular performance. Normal drivers may fear for vehicular instability, and it is thus difficult to brake and turn at the same time. Typical criteria for longitudinal and lateral acceleration parameters for proposed cautious, normal, aggressive, and extremely aggressive drivers, are plotted on the G–G diagram, shown in Fig. 2. 

\section{Occupant’s Preference Metric-aware Control System}

Fig. 3 describes an outline of the proposed system for occupant's preference-aware autonomous vehicle. Through the proposed system, a personalized autonomous driving service can be provided to car passengers. If we assume that an occupant of the fully autonomous self-driving vehicle can select his/her driving preference, the favorite driving style can be interpreted as a preference metric that the occupant might want for experience during riding a car. Firstly, at the preference metric selection stage, the occupant or passenger selects a preference-range of accelerations and jerks based on the proposed OPM. Once the time optimal velocity is calculated, this velocity profile is used for the reference input for controlling the velocity at the global path. This structure has the advantage of being able to correspond to the preference parameters for the motion planning while driving. On the curved road, the longitudinal velocity has a significant role because the value of the lateral acceleration largely depends on the longitudinal velocity of the vehicle.

The integrated longitudinal and the lateral controller is required to track the planned trajectory with the desired speed for various driving states of the autonomous vehicle. The proposed integrated vehicular control system generates steering control and throttle/brake maneuver inputs for the vehicles in order to perform simultaneous path and speed tracking, while ensuring the dynamic stability and tracking performance of the autonomous vehicle. More details on the integrated longitudinal and lateral control system, can be invoked from refs [7-8]. 

To satisfy the lateral acceleration preference, it is necessary to plan the cornering velocity according to the curvature of the road with respect to the lateral acceleration preference. For the appropriate velocity planning, we propose a time optimization method that satisfies the OPM constraint when the reference path is given from the starting point to the destination. The optimization problem is a well- known issue in robotics, to find optimal solution that minimizes the user defined cost of all possible solutions within constraints. The minimum time approach is one of the widely used methods for optimization problems. The minimum time optimization can be obtained by discretizing the continuous time model using the interior-point- method (IPM) based simultaneous approach [9]. We defined a nonlinear problem (NLP) to minimize the total travel time through the summation of the discrete time element duration. We propose a lane change path generation algorithm that satisfies the OPM condition while maintaining the current speed of vehicle since the sudden change of speed can cause discomfort to occupant. Fig. 4 shows the algorithm for proposed lane change path generation. First, if the lane change command is transmitted, it should be confirmed whether the lane change is possible or not. If the lane change maneuver is possible, the path planner selects a target point for the lane change and generates a continuous-curvature path. The path generation method and the target point selection had been studied in the previous works [7, 10].

\begin{figure}
        \centering
        \includegraphics[width=0.7\textwidth]{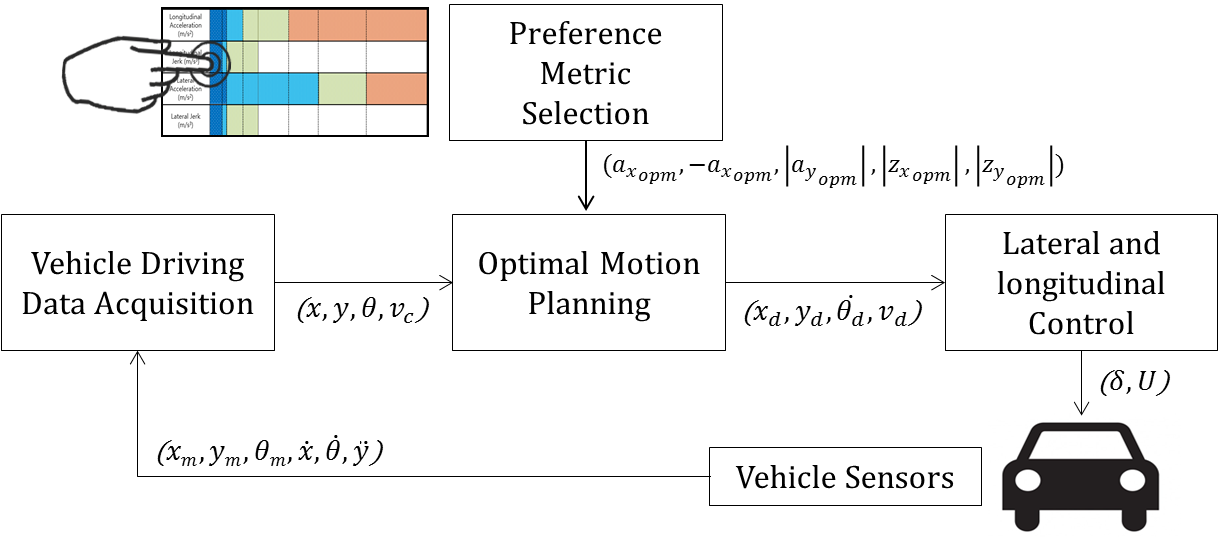}
        \caption{Outline of the proposed system architecture for user preference-aware autonomous vehicles.}
\end{figure}

\begin{figure}
        \centering
        \includegraphics[width=0.3\textwidth]{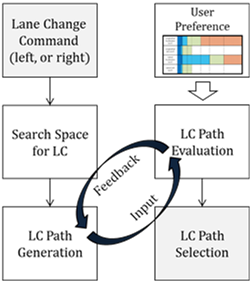}
        \caption{Proposed path generation and lane change algorithm.}
\end{figure}

\section{Simulation and Experimental Results}

To evaluate the feasibility of the proposed control system, we simulated the autonomous vehicle controller using Simulink and Carsim, and then implemented an experimental autonomous vehicle. Overall structure of the proposed upper-level control system is simple, easy to implement, and can be effectively applied to autonomous vehicles.

The vehicular parameters used for simulations are listed in Table 1. In the simulation we used a bicycle steering model for the vehicle. The vehicle mass and the wheelbase were $1,740kg$ and $3.05m$, respectively. In addition, the controllable front steering angle was limited within $\pm32^\circ$ considering the nonholonomic constraints of the vehicle. All acquired data were synchronized at the frequency of $10Hz$ during simulations.

\begin{table}
  \caption{Vehicular parameters used in the simulation.}
  \label{sample-table}
  \centering
  \begin{tabular}{lll}
    \toprule
    \cmidrule(r){1-3}
    Parameters     & Notation     & Value \\
    \midrule
    Vehicle mass & $m$ & $1740 kg$     \\
    Vehicle yaw moment     & $I_z$ & $3000 kg{\cdot}m^2$      \\
    Front-c.g. distance     & $l_f$       & $1.4 m$  \\
    Rear-c.g. distance     & $l_r$       & $1.65 m$  \\
    Cornering stiffness of front tires     & $C_{af}$       & $81000 N/\circ$  \\
    Cornering stiffness of rear tires     & $C_{ar}$       & $81000 N/\circ$  \\
    \bottomrule
  \end{tabular}
\end{table}

We conducted simulations for two different types of occupants, one with a cautious driving style, and the other for a dynamic driving style, with the following respective preferences,

\begin{equation}
OPM\#1=\left\{0.9,0.9,\left|0.9\right|,\left|0.6\right|,\left|0.6\right|\right\}
\end{equation}

\begin{equation}
OPM\#2=\left\{2.2,-2.5,\left|3.5\right|,\left|1.5\right|,\left|1.5\right|\right\}
\end{equation}

Fig. 6 and Fig. 7 display the simulated results of the velocity profile, lateral distance error during the route driving, longitudinal and lateral accelerations of the vehicle, and longitudinal and lateral accelerations on the G–G diagram for the reference route shown in Fig. 5. There is a lag in the simulated velocity compared to the calculated velocity. The longitudinal and lateral control block generated the thread maneuver that is input to the vehicle, and was used by the trajectory planning block to calculate the velocity to meet the OPM criterion. The delay of the vehicle’s response to the thread command caused a temporal delay between the calculated and the simulated values.

The results show that the lateral distance error does not exceed $5cm$, as shown in Fig. 6 (b) and Fig. 7 (b), both for cautious and aggressive driving conditions. The simulated accelerations and jerks could also be bounded within the OPM constraints. Depending on the OPM criterion, the difference of lap time is $122s$ between the cautious and the aggressive driving schemes. While the aggressive driving lasted for $145s$, the smooth comfort driving period lasted approximately $274s$.

As shown in Fig. 6 (c) and Fig. 7 (c), the driving pattern of the simulated self-driving vehicle yields a driving style that is similar to the heuristic pattern of deceleration-steering-acceleration. Human drivers normally decelerate when they approach a curved road. The simulated longitudinal and lateral acceleration profiles displayed on the G–G diagram show that some points are outside the borderline set by the criterion. In the case of the driving simulation for the OPM \#1, $88.3\%$ of the sampled data were inside the OPM acceleration constraints in the case of cautious driving, while in the case of OPM \#2, approximately $81.4\%$ of simulated data satisfied the OPM criterion. Even though most of the simulated longitudinal velocity did not surpass the boundary value, some of the lateral acceleration values were located outside the margins set by the criterion. The delay in the response of the vehicle’s velocity was the main reason for which the data points were outside the borderline, especially for the cases when the vehicle approached the curvy sections of the route. To improve the performance of the system, a faster response time for the overall control system is required. Although the proposed system could not completely control the vehicle to drive in compliance with the set OPM criterion throughout the entire driving path, the values that were outside the border were not far from the set criterion margins.

\begin{figure}
        \centering
        \includegraphics[width=0.6\textwidth]{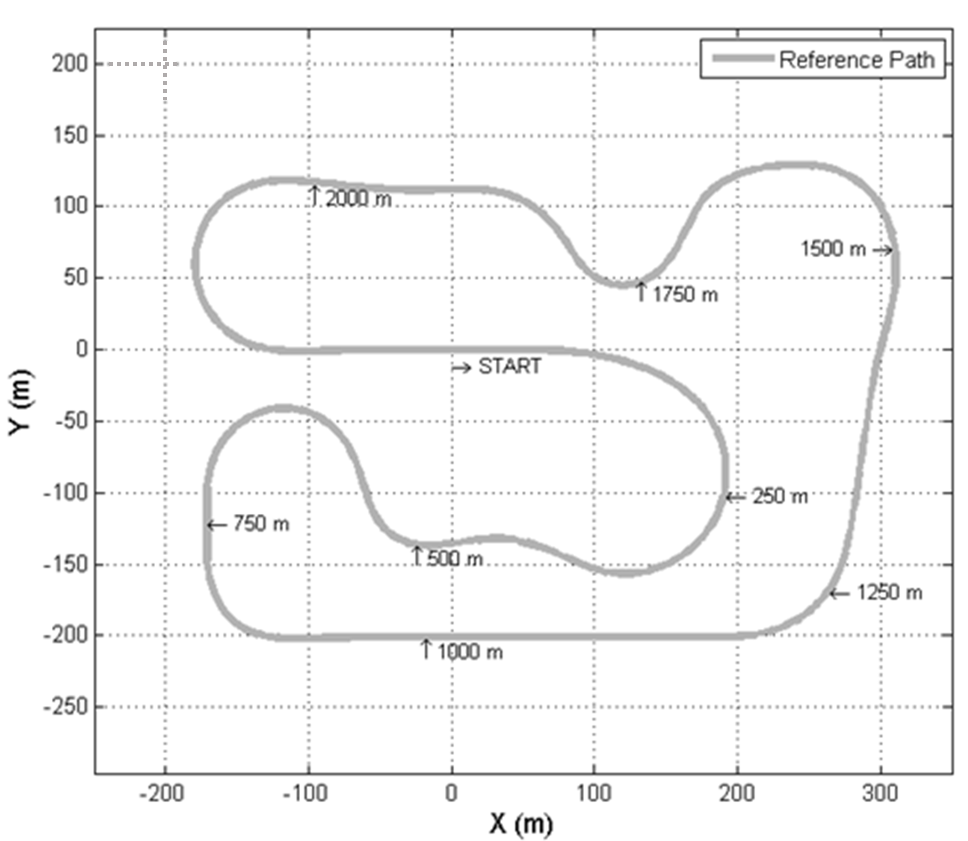}
        \caption{Reference path for the simulation.}
\end{figure}

\begin{figure}
        \centering
        \includegraphics[width=0.95\textwidth]{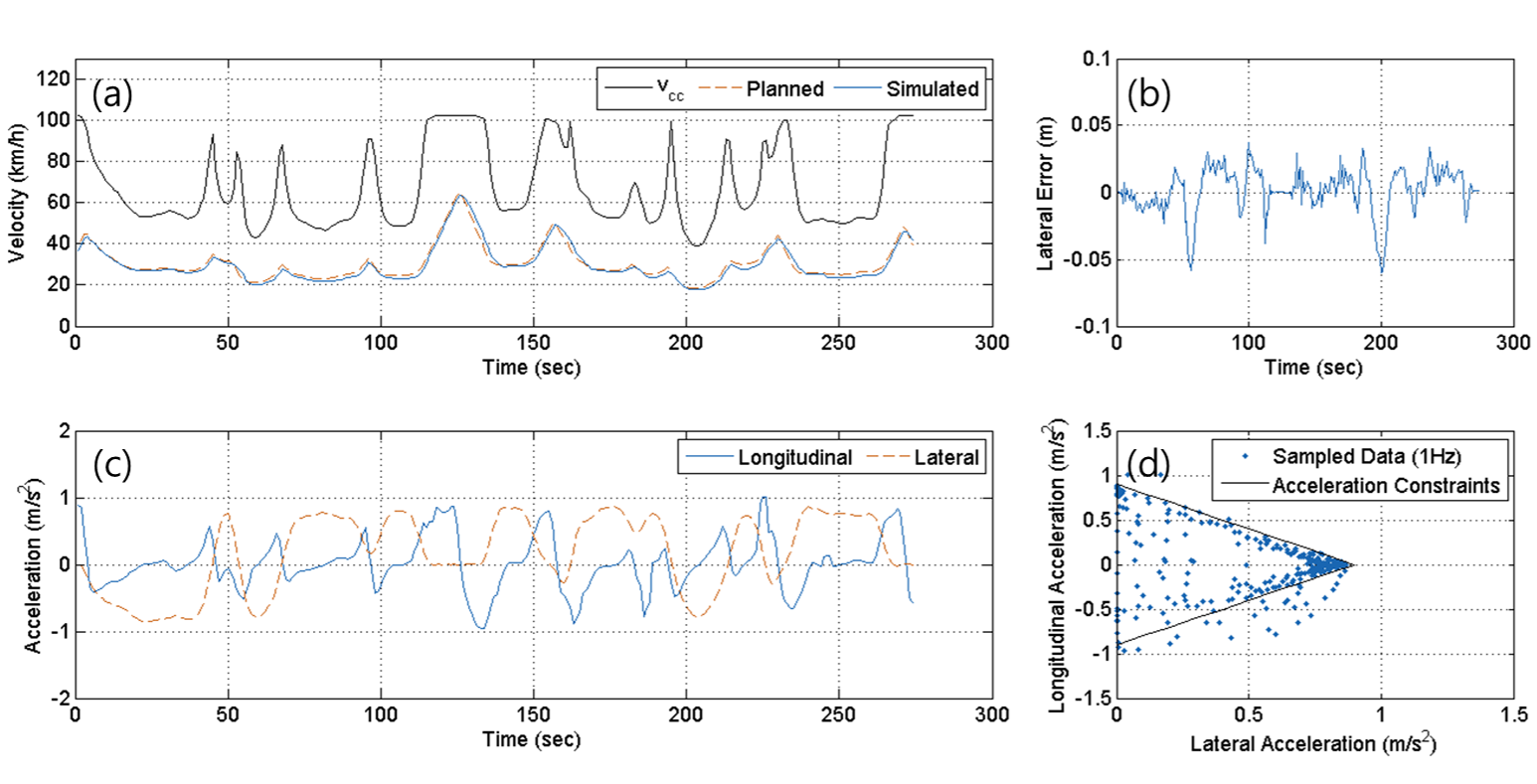}
        \caption{Simulated results of the OPM criterion $(OPM\#1={+0.9, -0.9, |0.9|, |0.6|, |0.6|})$. (a) Planned and simulated velocity, (b) simulated lateral distance error during driving, (c) simulated longitudinal and lateral acceleration, and (d) simulated longitudinal and lateral acceleration displayed on the G–G diagram.}
\end{figure}

\begin{figure}
        \centering
        \includegraphics[width=0.95\textwidth]{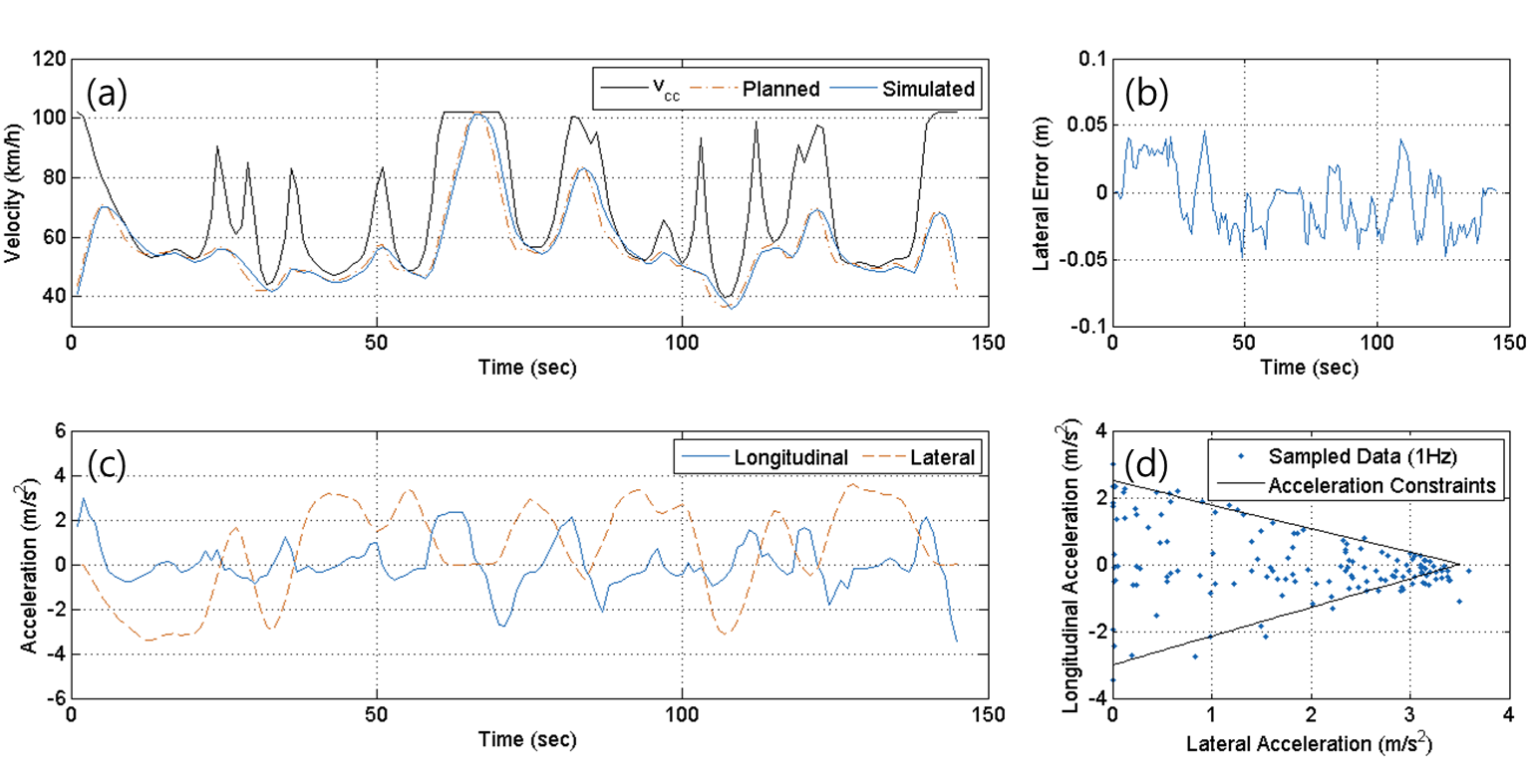}
        \caption{Simulated results of a typical aggressive driver’s OPM criterion $(OPM\#1={+2.2, -2.5, |3.5|, |1.5|, |1.5|})$. (a) Planned and simulated velocity, (b) simulated lateral distance error during driving, (c) simulated longitudinal and lateral acceleration, and (d) simulated longitudinal and lateral acceleration displayed on the G–G diagram.}
\end{figure}

\begin{figure}
        \centering
        \includegraphics[width=0.9\textwidth]{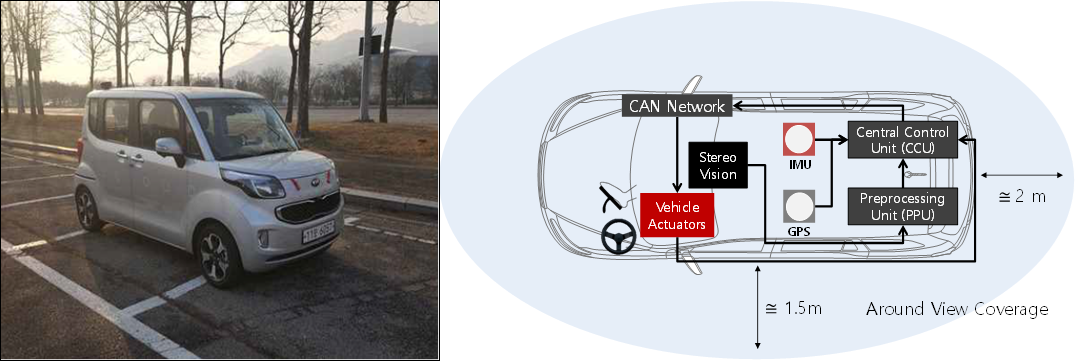}
        \caption{Photograph of the autonomous vehicle used for real-road experiments.}
\end{figure}

For the experimental test of the proposed control system on a real road, we modified the commercial vehicle RAY from Kia motors to an autonomous vehicle that was equipped with sensors, actuators, and the proposed controller, as shown in Fig. 8. In spite of the compactness of the car, the modified test vehicle provided enough space for all on-board equipment. This vehicle has a gasoline-powered engine, front-wheel drive, and motor-driven power steering (MDPS). The control commands of the steering angle and throttle maneuver to the vehicle were input to the vehicle through the CAN gateway network to control the vehicular actuators, such as the MDPS, and the throttle value. There was a separated actuator interface, but it was controlled by the CCU for maneuvering the gear shift lever and pressing brake pedal.

\begin{figure}
        \centering
        \includegraphics[width=0.9\textwidth]{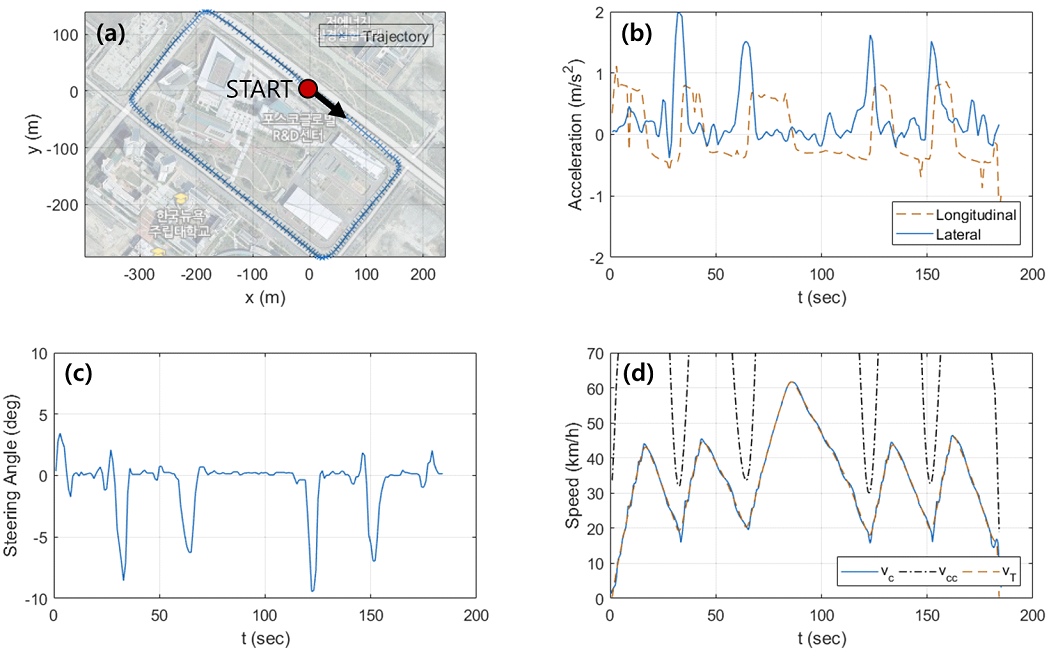}
        \caption{Experimental results in a case of OPM criterion for $(OPM ={+0.6, -0.6, |1.5|, |0.6|, |0.6|})$. (a) Satellite map of the test route, (b) measured accelerations of the vehicle, (c) measured front-steering angle, (d) velocity profile obtained from in-vehicle CAN during real-road tests.}
\end{figure}

We conducted experiments on the real-road course, where, the testing course was $1.3km$ long and which is useful to evaluate the performance of the proposed controller because of 4 turning course with/without traffic signals. In the real-road test, we set the OPM criterion to match the various passenger driving styles. Fig. 9 shows the measured results of the field test for autonomous driving experiments in a case of specific OPM criterion for the typical normal driving style, $(OPM =\left\{+0.6, -0.6, |1.5|, |0.6|, |0.6|\right\})$. The acceleration data were acquired by the IMU sensor, and velocity and steering angle data were obtained through a CAN interface during the test. Measured experimental data illustrate that vehicular velocity followed the planed velocity during driving. The measured acceleration and velocity profiles of the self-driving vehicle indicate a driving style that is similar to the heuristic pattern of deceleration-steering-acceleration. Fig. 9 (b) shows the recorded longitudinal and lateral accelerating profiles, where the vehicle decelerates while it approaches an intersection, and makes a turn before it accelerates again.

The measured lateral acceleration profile shows four overshoots coincident with the events associated with changing steering angles for right turns at intersections. We found that the measured values for lateral acceleration at the overshooting points were equal to $1.93 m/s^2$ for the first turn, and ranged from $1.5$ to $1.6m/s^2$ for the other turns. The measured value of the lateral acceleration seemed to be limited, but the values exceeded the given OPM criterion during turning maneuvers. Additionally, there were glitches in the measured longitudinal acceleration, but the overall values were regulated not to grow beyond the given criterion. This measured trend of the longitudinal and lateral acceleration values at the curved section of the route was consistent with the simulated results since only the lateral acceleration results slightly exceeded the criterion limit, while the longitudinal acceleration was within the set OPM boundary. Owing to a latency in the vehicle’s velocity response with respect to the desired input velocity, the vehicular speed during turning was slightly higher than the planned velocity required to meet the criterion of lateral acceleration. This might be the main reason for which the lateral acceleration value was marginally outside the border line.

The OPM criterion included lateral and longitudinal jerks. However, we could not measure jerk in the experiments using the IMU sensor because of the measurement noise in the acceleration data acquisition. Jerk is the derivative of acceleration, and even low-noise amplitudes in the acceleration data can cause spikes in the estimation of jerk owing to the temporal derivative of the data. Although various studies have been conducted for the improvement of jerk sensor, to the best of our knowledge, on-vehicle lateral or longitudinal jerk sensors for robust data acquisitions have not been commercialized thus far. Therefore, in the proposed controller, the jerk term can be taken into account only at the stage of the planning of the desired velocity.

\section{Discussions and Concluding Remarks}

In this work, we proposed the occupant’s preference-aware control system for the personalized autonomous driving service. For this purpose, we designed the occupant preference metric defining a preferred lateral and longitudinal accretion region with the maximum allowable jerk for the autonomous vehicle users. The proposed controller is based on the motion control parameters enabling integrated lateral and longitudinal control via preference-aware maneuvering of autonomous vehicles. The overall control system and whole planning strategies are verified using the real autonomous driving vehicle in the urban roadway including lane change maneuvers. The simulations and experimental results of the implemented OPM-aware controller demonstrated that the proposed system controlled the self-driving vehicle according to the specified criterion of admissible acceleration and jerk.

%

\section*{References}

\small

[1] M. Elbanhawi, M. Simic, and R. Jazar. "In the passenger seat: investigating ride comfort measures in autonomous cars," IEEE Intelligent Transportation Systems Magazine 7(3), 4–17, (2015)

[2] M. Turner, and G. Michael, "Motion sickness in public road transport: the effect of driver, route and vehicle," Ergonomics 42(12), 1646–1664, (1999)

[3] D. Martin, and D. Litwhiler, "An Investigation of Acceleration and Jerk Profiles of Public Transportation Vehicles," presented at the American Society for Engineering Education, AC 2008-1330, (2008)

[4] S. Moon, and K. Yi, "Human driving data-based design of a vehicle adaptive cruise control algorithm," Vehicle System Dynamics 46(8), 661–690, (2008)

[5] P. Bosetti, L. Mauro, and S. Andrea, "On curve negotiation: From driver support to automation," IEEE Transactions on Intelligent Transportation Systems, vol. 16, no. 4, pp. 2082-2093, (2015)

[6] P. Bosetti, L. Mauro, and S. Andrea, "On the human control of vehicles: an experimental study of acceleration," European Transport Research Review, vol. 6, no. 2, pp. 157-170, (2014)

[7] I. Bae, J. Kim, J. Moon, and S. Kim, "Lane Change Maneuver based on Bezier Curve providing Comfort Experience for Autonomous Vehicle Users," , IEEE International Conference on Intelligent Transportation Systems (ITSC), October (2019)

[8] I. Bae, J. Moon, and J. Seo, “Toward a Comfortable Driving Experience for a Self-Driving Shuttle Bus,” pp. No. 8, Electronics (2019)

[9] B. Li, and S. Zhijiang, "Simultaneous dynamic optimization: A trajectory planning method for nonholonomic car-like robots," Advances in Engineering Software, pp. 30-42, (2015)

[10] I Bae, J Moon, S Kim, "Driving Preference Metric-Aware Control for Self-Driving Vehicles", International Journal of Intelligent Engineering and Systems 12 (6), 157-166, (2019)

\end{document}